\documentclass{aipproc}
\begin{document}

\title{Algorithm for Computing Excited States in Quantum Theory}

\author{X.Q. Luo$^1 \footnote{Email: stslxq@zsu.edu.cn}$,
H. Jirari$^2$, H. Kr\"oger$^2$, and K. Moriarty$^3$}

\affiliation{$^1$Department of Physics, Zhongshan University, 
Guangzhou 510275, China\\
$^2$D\'{e}partement de Physique, Universit\'{e} Laval, Qu\'{e}bec,
  Qu\'{e}bec G1K 7P4, Canada \\
$^3$Department of Mathematics, Statistics and Computer Science,
  Dalhousie University, Halifax, Nova Scotia B3H 3J5, Canada}

\begin{abstract}
Monte Carlo techniques have been widely employed in statistical physics
as well as in quantum theory in the Lagrangian formulation.
However, in the conventional approach,
it is extremely
difficult to compute the excited states.
Here we present a different algorithm: the Monte Carlo Hamiltonian
method,
designed to overcome the difficulties of the conventional approach.
As a new example, application to the Klein-Gordon field theory is shown.
\end{abstract}

\maketitle

\section{Introduction}

There are two standard formulations in quantum theory:
Hamiltonian and Lagrangian.
A comparison of the conventional approaches is given in Tab. 1.

Monte Carlo (MC) method with importance sampling is an excellent
non-perturbative technique to calculate path integrals in quantum theory.
In the last two decades, and it has successfully been applied 
to Lagrangian lattice gauge theory \cite{Creutz,Rothe,Montvay}.
In the standard Lagrangian MC method, however,
it is extremely difficult to compute the spectrum 
and wave function beyond the ground state.
On the other hand, 
the standard Hamiltonian formulation is capable of
doing it. 
 
Recently, we proposed an algorithm to construct an effective
Hamiltonian from Lagrangian MC simulations in Ref. \cite{mch}.
We called it the MC Hamiltonian method.
The advantage, in comparison with the standard Lagrangian MC approach, 
is that one can obtain the spectrum and wave functions beyond
the ground state. It also allows to do thermodynamics. 
In this paper, we briefly review what we have done, and
present some new results.

\begin{table}
\begin{tabular}{|c|c|c|}
\hline
{\bf  Formulation} & {\bf Hamiltonian}  &  {\bf Lagrangian} \\
\hline
Approach   & Schr\"odinger Eq.  &  Path Integral\\
           & $H \vert E_n> = E_n \vert E_n>$
                                            &  $<O> =$\\
           &                                &     ${\int d[x] O[x]
                                                   \exp(- S[x]/ \hbar)  \over
                                                    \int [dx] \exp(- S[x]/ \hbar)}$ \\
\hline
Algorithm & Series expansion,               & MC simulation      \\
          & variational approx.,            & with importance    \\
          & Runge-Kutta ...                 & sampling           \\
\hline
Advantage  & Both ground state,             & It generates        \\
           & $\&$ excited states            & the important       \\ 
           & can be computed.               & configurations      \\
\hline
Problem  & Difficult for                    & Difficult for \\
         & many body systems.                & excited states.  \\
\hline            
\end{tabular}
\caption{Comparison of the conventional methods.}
\end{table}


\section{Algorithm}

\subsection{Effective Hamiltonian}
Let us review briefly the basic ideas of our approach.  
The (imaginary time) transition amplitude between 
an initial state at position  
$x_i$, and time $t_i$,  
and final state at $x_f$ and $t_f$ is related 
to the Hamiltonian $H$ by
\begin{eqnarray*}
M_{fi}=<x_{f},t_{f} | x_{i},t_{i}> 
= <x_{f} | e^{-H(t_{f}-t_{i})/\hbar} | x_{i}>
\end{eqnarray*}
\begin{eqnarray}
= \sum_{n=1}^{\infty} <x_{f} | E_{n} > 
e^{-E_{n} T/\hbar} < E_{n} | x_{i}>,
\end{eqnarray}
where $T=t_f-t_i$.
According to Feynman's path integral formulation of quantum mechanics (Q.M.),
the transition amplitude is also related to the path integral:
\begin{eqnarray}
M_{fi} =
\int [dx] \exp(- S[x]/ \hbar) \vert_{x_i,t_i}^{x_f,t_f}.
\end{eqnarray}
The starting point of our method, as described in detail in Ref. \cite{mch}
is to construct an effective Hamiltonian $H_{eff}$ 
(finite $N \times N$ matrix) by 
\begin{eqnarray*}
M_{fi} =
<x_{f} | e^{-H_{eff} T/\hbar} | x_{i}>
\end{eqnarray*}
\begin{eqnarray}
=\sum_{n=1}^{N} < x_{f} | E^{eff}_{n}> e^{-E^{eff}_{n} T/\hbar} 
< E^{eff}_{n} | x_{i} > .
\end{eqnarray}
The eigenvalues $E^{eff}_{n}$ and wave function
$\vert E^{eff}_{n} >$  can be obtained,
by diagonalizing $M$ using a unitary transformation
\begin{eqnarray}
M = U^{\dagger}DU,
\end{eqnarray}
where $D =diag (e^{-E^{eff}_{1}T/\hbar},..., e^{-E^{eff}_{N}T/\hbar})$.
Once the spectrum
and wave functions are available, 
all physical information can also be obtained.

Since the theory described by $H$, which basis in Hilbert space is infinite, 
is now approximated by a theory
described by a finite matrix $H_{eff}$, which basis is finite,
the physics
of $H$ and $H_{eff}$ might be quite different at high energy. 
Therefore we expect that we can only
reproduce the low energy physics of the system.
This is good enough for our purpose. 
In Refs. \cite{mch,physica,lat99_1,Jiang}, we investigated many 1-D, 2-D and 3-D Q.M. 
models (Tab. 2) using this MC Hamiltonian algorithm.
We computed the spectrum, wave functions and some thermodynamical observables.
The results are in very good agreement with those from analytical and/or Runge-Kutta methods.

\begin{table}[hbt]
\caption{Q.M. systems, investigated by the MC Hamiltonian method using the regular basis.}
\begin{tabular}{|c|c|c|}
\hline
System & Potential\\
\hline

Q.M. in 1-D	& $V(x)=0$\\
                & $V(x)={1\over 2} m \omega^2 x^2$\\ 
        	& $V(x)=-V_0 \rm{sech}^2(x)$\\
        	& $V(x)={1\over 2} x^2+{1\over 4} x^4$\\
        	& $V(x)={1 \over 2} \vert x \vert$\\
                & $V(x) =\{
                       \begin{array}{cc}
                         \infty,        & x < 0 \\
                         Fx,            &  x \ge  0 
                  \end{array}$\\ 
\hline
Q.M. in 2-D   & $V(x,y)={1\over 2} m \omega^2 x^2+{1\over 2} m \omega^2 y^2$\\
              & $V(x,y)={1\over 2} m \omega^2 x^2+{1\over 2} m \omega^2 y^2+\lambda xy$\\
\hline
Q.M. in 3-D  & $V(x,y,z)={1\over 2} m \omega^2 x^2+{1\over 2} m \omega^2 y^2+{1\over 2} m \omega^2 z^2$\\
\hline
\end{tabular}
\end{table}

\subsection{Basis in Hilbert Space}

To get the correct scale for the spectrum, the 
position state $\vert x_n \rangle$  (Bargman states or box states) 
at $t_i$ or $t_f$ should be properly normalized. We denote a normalized basis 
of Hilbert states as 
$\vert e_n \rangle$, $n=1, ..., N$. In position space, it can be
expressed as
\begin{eqnarray}
e_n(x) =\{
\begin{array}{cc}
     1/\sqrt{\Delta x_n},        & x \in [x_n,x_{n+1}] \\
     0,                         &  x \notin  [x_n,x_{n+1}] 
\end{array}
\end{eqnarray}
where $\Delta x_n=x_{n+1}-x_n$.
 
The simplest choice is a basis with $\Delta x_n =const.$, which is called the `` regular basis''.
In Refs. \cite{mch,physica,lat99_1,Jiang}, the regular basis is used.
For many body system or quantum field theory, the regular basis will encounter problem.
For example, in a system with a 1-D chain of oscillators (see later), if the number of oscillators
is 30, the minimum non-trivial regular basis is $N=2^{30}=1073741824$, which is prohibitively large
for numerical calculations.

Guided by the idea of important sampling, in Refs. \cite{lat99_2,Huang}, we proposed to select a basis
from the Boltzmann weight proportional to the transition amplitude between $x'_i=0$ at $t'_i=0$ and $x'_f=x_n$
at some $t'_f$.
In a free particle or harmonic oscillator case, the distribution is just a Gaussian
\begin{eqnarray}
P_{basis}[x_n] ={1 \over \sqrt{2\pi} \sigma} \exp(-{x_n^2 \over 2 \sigma^2}),
\label{STO}
\end{eqnarray}
where $\sigma=\sqrt{\hbar t'_f/m}$ for the free case 
and  $\sigma=\sqrt{\hbar {\rm sinh}(\omega t'_f)/(m \omega)}$ 
for the harmonic oscillator. 
We call such a basis the ``stochastic basis''.

\subsection{Matrix elements}
As explained above, 
the calculation of the transition matrix elements is an essential ingredient of our method.
The matrix element in the normalized basis is related to 
$<x_{n'},t_{f} | x_{n},t_{i}>$ by 
\begin{eqnarray}
M_{n'n}
&=& 
<e_{n'},T \vert e_{n},0>
\nonumber \\
&=& 
\int_{x_n'}^{x_{n'+1}} dx' \int_{x_n}^{x_{n+1}} dx'' 
{ <x',t_{f} | x'',t_{i}> \over \sqrt{\Delta x_{n'}\Delta x_{n}}}
\nonumber \\
&\approx& 
\sqrt{\Delta x_{n'} \Delta x_{n}} <x_{n'},t_{f} | x_{n},t_{i}>,
\label{ME}
\end{eqnarray}
where $<x_{n'},t_{f} | x_{n},t_{i}>$ can be calculated using MC simulations as follows.

\noindent
(a) Discretize the continuous time.

\noindent
(b) Generate free configurations $[x]$ 
between $t \in (t_i,t_f)$ obeying the Boltzmann distribution
\begin{eqnarray}
P_0[x] = { \exp(- S_0[x]/ \hbar)  \over
       \int [dx] \exp(- S_0[x]/ \hbar)}\vert_{x_n,t_i}^{x_{n'},t_f},
\end{eqnarray}
where $S_0=\int dt ~ m\dot{x}^2/2$.

\noindent
(c) Measure
\begin{eqnarray} 
<O_V>=
\int [dx] \exp(- \int dt ~ V(x)/ \hbar)\vert_{x_n,t_i}^{x_{n'},t_f} P_0[x]
\end{eqnarray}
The path integral in Eq. (\ref{ME}) is then
\begin{eqnarray}
<x_{n'},t_{f} | x_{n},t_{i}> &=& <O_V> \times \sqrt{m \over 2\pi \hbar T}
\nonumber \\
& \times & 
 \exp [ -{m \over 2 \hbar T} (x_{n'}-x_n)^2 ].
\end{eqnarray}

\section {Quantum Field Theory}   

The main purpose of the algorithm is to study
many body systems and quantum field theory beyond the ground state.
As an example, we consider
a chain of $N_{osc}$ coupled oscillators in 1 spatial dimension. 
Its Hamiltonian is given in Ref. \cite{Henley} as
\begin{eqnarray}
H=\sum_{j=1}^{N_{osc}} {1 \over 2} [p_j^2+\Omega^2 (q_j-q_{j+1})^2+\Omega_0^2 q_j^2],
\end{eqnarray}
where $p_j$ and $q_j$ are the momentum and displacement of the j-th oscillator respectively.
This model is equivalent to the Klein-Gordon field theory on a (1+1)-dimensional lattice.

The spectrum of the system is analytically known:  
\begin{eqnarray}
E_{n}&=&\sum_{n_k} ~~ (n_k+{1\over 2})  \hbar \omega_k, 
\nonumber \\
\omega_k&=&\sqrt{\Omega^2(2 \sin k/2)^2 +\Omega_0^2}, 
\end{eqnarray}
where $n_1,...,n_{N_{osc}}= 0,1,\ldots$, and $k=2 \pi l/N_{osc}$ with $l$ 
an integer between $-N_{osc}/2$ and $N_{osc}/2$.

We generate a stochastic basis according to Eq. (\ref{STO})
with N=1000 configurations $[q_1,...,q_{N_{osc}}]$ for the
initial and final states for  $N_{osc}=9$ oscillators. For the adjustable parameter $\sigma$, we use the distribution for the uncoupled oscillators 
at $t'_f=t_f$
for simplicity. (Of course, one should study systematically the dependence of the results on $\sigma$). 
Tab. 3 compares the spectrum from the MC Hamiltonian  
with the analytical results for the first 20 states with $\Omega=1$, $\Omega_0=2$, $m=1$, $\hbar=1$, and $T=2$.
They agree very well with the exact ones.

\begin{table}[hbt]
\caption{Comparison of the spectrum of the chain of 9 coupled oscillators,
between the MC Hamiltonian method with a {\it stochastic} basis and the analytic ones.}
\begin{tabular}{|c|c|c|}
\hline
$n$ & $E_{n}^{eff}$  & $E_{n}^{Exact}$\\
\hline
   1   & 10.904663192168  &  10.944060480668\\
   2   & 12.956830557334  &  12.944060480668\\
   3   & 12.985023578737  &  13.057803869484\\
   4   & 13.044311582647  &  13.057803869484\\
   5   & 13.299967341242  &  13.321601993380\\
   6   & 13.345480638394  &  13.321601993380\\
   7   & 13.552195133687  &  13.589811791733\\
   8   & 13.585794986361  &  13.589811791733\\
   9   & 13.680136748933  &  13.751084748745\\
   10  &  13.744919087477 &   13.751084748745\\
   11  &  14.984737011385 &   14.944060480668\\
   12  &  15.012353803145 &   15.057803869484\\
   13  &  15.057295761044 &   15.057803869484\\
   14  &  15.108904652020 &   15.171547258300\\
   15  &  15.125356713561 &   15.171547258300\\
   16  &  15.187413290039 &   15.171547258300\\
   17  &  15.308536490102 &   15.321601993380\\
   18  &  15.396255686587 &   15.321601993380\\
   19  &  15.420708031412 &   15.435345382196\\
   20  &  15.432823810789 &   15.435345382196\\
\hline
\end{tabular}
\end{table}

\section{Summary}

In this paper, we have tested the MC Hamiltonian method with a stochastic basis in
a many body Q.M. system with a chain of coupled oscillators: 
the Klein-Gordon field theory on a (1+1)-dimensional lattice.
The results are very encouraging.
We believe that the application of the algorithm to more complicated body systems and quantum field theory
will be very interesting.

\section{Acknowledgements}
X.Q.L. is supported by the
National Science Fund for Distinguished Young Scholars (19825117),
National Science Foundation, 
Guangdong Provincial Natural Science Foundation (990212) and 
Ministry of Education of China.
H.K. and K.M. is supported by NSERC Canada.

\bibliographystyle{aipproc}

\end{document}